\documentclass[twocolumn,twoside,slac_two]{revtex4}
\usepackage{graphicx}
\usepackage{fancyhdr}
\begin{document}

\title{Soft-gluon and hadron-mass effects on fragmentation functions}

%

\author{B. A. Kniehl}
\affiliation{{II.} Institut f\"ur Theoretische Physik, Universit\"at Hamburg,
Luruper Chaussee 149, 22761 Hamburg, Germany}

\begin{abstract}
We review recent progress in the development of an approach valid to any order
which unifies the fixed-order DGLAP evolution of fragmentation functions at
large $x$ with soft-gluon logarithmic resummation at small $x$.
At leading order, this approach, implemented with the Double Logarithmic 
Approximation, reproduces exactly the Modified Leading Logarithm
Approximation, but it is more complete due to the degrees of freedom given to
the quark sector and the inclusion of the fixed-order terms.
We find that data from the largest $x$ values to the peak region can be better
fitted than with other approaches.
In addition, we develop a treatment of hadron mass effects that leads to
additional improvements at small $x$.
\end{abstract}

\maketitle

\thispagestyle{fancy}


\section{Introduction}
\label{sec:one}

The perturbative approach to Quantum Chromodynamics (QCD) is believed
to solve all problems within its own limitations provided the correct
choices of the expansion variable and the variable(s) to be fixed are
used. However, perturbative QCD (pQCD) currently has the status of
being a large collection of seemingly independent approaches, since a
single unified approach valid for all processes is not known. This is
problematic when one wants to use a range, qualitatively speaking, of
different processes to constrain the same parameters, {\it e.g.}\ 
in global fits. What is needed is a single formalism valid over the union of
all ranges that the various pQCD approaches allow. This unification must be
consistent, {\it i.e.}\ it must agree with each approach
in the set, when the expansion of that approach is used up to the
order being considered.

In this presentation, we review recent progress in establishing a generalized
formalism for the evolution of fragmentation functions (FFs) from the smallest
$x$ values probed so far way up to unity \cite{Albino:2005gg}.
After explaining how to treat soft-gluon and hadron-mass effects in
Secs.~\ref{sec:two} and \ref{sec:three}, respectively, we study their
phenomenological implications in Sec.~\ref{sec:four}.
Our conclusions are contained in Sec.~\ref{sec:five}.

\section{Resummation of soft-gluon logarithms}
\label{sec:two}

The current optimum description of single-hadron inclusive production is
provided by the QCD parton model, which requires knowledge of the FFs
$D_a^h(x,Q^2)$, each corresponding at leading order (LO) to the probability
for the parton $a$ produced at short distance $1/Q$ to form a jet that
includes the hadron $h$ carrying a fraction $x$ of the longitudinal momentum
of $a$.
From now on we omit the label $h$ and put the FFs for all $a$ into the vector
$D(x,Q^2)$.
The evolution of the FFs in the factorization scale $Q^2$ is also required,
which at large and intermediate $x$ is well described \cite{KKP2000} by
the Dokshitzer-Gribov-Lipatov-Altarelli-Parisi (DGLAP) equation \cite{DGLAP},
given at LO by
\begin{equation}
\frac{d}{d\ln Q^2} D(x,Q^2)=\int_x^1 \frac{dz}{z}a_s(Q^2) P^{(0)}(z) D
\left(\frac{x}{z},Q^2\right),
\label{DGLAPx}
\end{equation}
where $P^{(0)}(z)$ are the LO splitting functions calculated from fixed-order
(FO) pQCD. 
We define $a_s=\alpha_s/(2\pi)$, which at LO obeys $a_s(Q^2)={}$\break
$1/[\beta_0\ln(Q^2/\Lambda_{\rm QCD}^2)]$,
where $\beta_0=(11/6)C_A-(2/3)T_R n_f$ is the
first coefficient of the beta function and $\Lambda_{\rm QCD}$ is the
asymptotic scale parameter of QCD.
For the color gauge group SU(3), the color factors appearing in this
presentation are $C_F=4/3$, $C_A=3$, and $T_R=1/2$; $n_f$ is the number of
active quark flavors.
 
On the other hand, at small $x$, the FO approximation fails due to unresummed
{\it soft-gluon logarithms} (SGLs). The most singular 
SGLs are the {\it double logarithms} (DLs), which give a
$z\rightarrow 0$ singularity in the LO splitting function $a_s P^{(0)}(z)$ 
of the form $a_s/z$. These DLs occur at all orders in the FO
splitting function, being generally of the form
$(1/z) (a_s \ln z)^2 (a_s \ln^2 z)^r$ for
$r=-1,\ldots,\infty$, implying that,
as $x$ decreases, Eq.~(\ref{DGLAPx}) becomes a poor approximation
once $\ln (1/x) = O(a_s^{-1/2})$. 
An improvement of the small-$x$ description should be obtained by
accounting for all DLs, which is provided by
the Double Logarithmic Approximation
(DLA) \cite{Bassetto:1982ma;Fadin:1983aw,Dokshitzer:1991wu},
\begin{eqnarray}
\frac{d}{d \ln Q^2}D(x,Q^2)&=&\int_x^1 \frac{dz}{z}\, \frac{2C_A}{z}
A  z^{2\frac{d}{d\ln Q^2}}
\left[a_s(Q^2)
\vphantom{\left(\frac{x}{z},Q^2\right)}\right.\nonumber\\
&&{}\times\left.D\left(\frac{x}{z},Q^2\right)\right],
\label{DLAx}
\end{eqnarray}
where $A=0$ for valence-quark or non-singlet FFs, while 
\begin{equation}
A=\left( \begin{array}{cc}
0 & \frac{2 C_F}{C_A} \\
0 & 1
\end{array} \right),
\end{equation}
when $D=(D_{\Sigma},D_g)$, where $D_{\Sigma}=(1/n_f)\sum_{q=1}^{n_f}
(D_q +D_{\overline{q}})$ is the singlet FF.
The Modified Leading Logarithm Approximation (MLLA) 
\cite{Dokshitzer:1991wu,Mueller:1982cq,Dokshitzer:1984dx} improves the
description over the DLA only by including a part of the FO contribution that
is known to be important at small $x$.
With certain qualifications \cite{Albino:2004yg}, the MLLA leads to a good
description of all data down to the smallest $x$ values.

However, what has been lacking is a single approach which can describe data
from the largest to the smallest $x$ values. 
Such an approach can be simply and consistently constructed from the knowledge
provided by the DLA and the DGLAP equation.
The result reproduces the FO result when expanded in $a_s$, while all
small-$\omega$ singularities are resummed into a non-singular expression in
Mellin space (see later).
As shown in detail in Ref.~\cite{Albino:2005gg}, this approach is unique up to
higher-order terms in the FO series and up to less singular SGLs,
{\it e.g.}\ the {\it single logarithms} (SLs). 
Such a formalism may be constructed simply by
evolving according to Eq.~(\ref{DGLAPx}), but with
the splitting function modified as
\begin{equation}
a_s P^{(0)}(z) \rightarrow P^{\rm DL}(z,a_s)
+a_s \overline{P}^{(0)}(z),
\label{replace}
\end{equation}
where $P^{\rm DL}(z,a_s)$ contains the complete resummed contribution to the
splitting function to all orders from the DLs, while
$a_s\overline{P}^{(0)}(z)$ is the remaining FO contribution at LO.
It is obtained
by subtracting the LO DLs, already accounted for in $P^{\rm DL}$,
from $a_s P^{(0)}(z)$ to prevent double counting.
We work to LO, since the less singular SGLs that occur at next-to-leading
order (NLO) are not known to all orders.
However, we leave the SL at LO unresummed, since it is not singular as
$z\rightarrow 0$.
We now use Eq.~(\ref{DLAx}) to gain some understanding of $P^{\rm DL}$.
For this, we transform to Mellin space, defined by
\begin{equation}
f(\omega)=\int_0^1 dx\, x^{\omega} f(x).
\label{Meltransdef}
\end{equation}
Upon Mellin transformation, Eq.~(\ref{DLAx}) becomes
\begin{eqnarray}
\lefteqn{\left(\omega+2\frac{d}{d \ln Q^2} \right) \frac{d}{d \ln Q^2}
D(\omega,Q^2)}
\nonumber\\
&=&2C_A a_s(Q^2)A D(\omega,Q^2).
\label{DRAPpre}
\end{eqnarray}
Substituting Eq.~(\ref{replace}) with the FO term $a_s\overline{P}^{(0)}(z)$
neglected in Eq.~(\ref{DGLAPx}), taking the Mellin transform,
\begin{equation}
\frac{d}{d\ln Q^2}D(\omega,Q^2)=P^{\rm DL}(\omega,a_s(Q^2))D(\omega,Q^2),
\label{DGLAPDLn}
\end{equation}
and inserting this in Eq.~(\ref{DRAPpre}), we obtain
\begin{equation}
2(P^{\rm DL})^2+\omega P^{\rm DL}-2C_A a_s A=0.
\label{DLAeqsimplest}
\end{equation}
We choose the solution
\begin{equation}
P^{\rm DL}(\omega,a_s)=\frac{A}{4}\left(-\omega+\sqrt{\omega^2+16C_A a_s}
\right),
\label{DLresummedinP}
\end{equation}
since its expansion in $a_s$ yields at LO the result
\begin{equation}
a_s P^{{\rm DL}(0)}(\omega,a_s)=
\left( \begin{array}{cc}
0 & a_s \frac{4 C_F}{\omega} \\
0 & a_s \frac{2 C_A}{\omega}
\end{array} \right),
\label{NLODLinmelspace}
\end{equation}
which agrees with the LO DLs from the literature \cite{DGLAP}.
Equation~(\ref{DLresummedinP}) contains all terms in the splitting function of
the form $(a_s/\omega)(a_s/\omega^2)^{r+1}$,
being the DLs in Mellin space, and agrees with the results of
Refs.~\cite{Dokshitzer:1991wu,Mueller:1982cq}.
We now return to $x$ space, where Eq.~(\ref{DLresummedinP}) reads
\begin{equation}
P^{\rm DL}(z,a_s)=\frac{A\sqrt{C_A a_s}}{z\ln(1/z)}
J_1\left(4\sqrt{C_A a_s}\ln \frac{1}{z}\right),
\label{allDLinzindelPclosed}
\end{equation}
with $J_1$ being the Bessel function of the first kind. 

To summarize our approach, we evolve the FFs according to
Eq.~(\ref{DGLAPx}), but with the replacement of Eq.~(\ref{replace}), where
$P^{\rm DL}(z,a_s)$ is given by Eq.~(\ref{allDLinzindelPclosed}), and
$a_s\overline{P}^{(0)}(z)$ is given by $a_sP^{(0)}(z)$ after the terms
proportional to $a_s/z$ have been subtracted.

To extend NLO calculations to small $x_p$, the complete resummed DL
contribution given by Eq.~(\ref{allDLinzindelPclosed}) must be added
to the NLO splitting functions.
These contain SGLs belonging to the classes $m=1,\ldots,4$, which must be
subtracted.
Note that the NLO $m=1$ term is accounted for by the resummed DL contribution.
The $m=4$ term is a type $p=0$ term, and hence does not need to be subtracted.

Before we outline the phenomenological investigation of our approach, we note
that it is more complete than the MLLA, which can be shown as follows.
With $a_s \overline{P}^{(0)}(z)$ accounted for, Eq.~(\ref{DRAPpre}) is
modified to
\begin{eqnarray}
\lefteqn{\left(\omega+2\frac{d}{d \ln Q^2} \right) \frac{d}{d \ln Q^2}
D(\omega,Q^2)}
\nonumber\\
&=&2C_A a_s(Q^2) A D(\omega,Q^2)
+\left(\omega+2\frac{d}{d \ln Q^2}\right)\left[a_s(Q^2)
\vphantom{\overline{P}^{(0)}}\right.
\nonumber\\
&&{}\times\left.\overline{P}^{(0)}(\omega)
D(\omega,Q^2)\right],
\label{DRAP}
\end{eqnarray}
up to terms which are being neglected in this presentation and which are
neglected in the MLLA. If we approximate $a_s\overline{P}^{(0)}(\omega)$
by its SLs, defined at LO to be the coefficients of $\omega^0$,
equal to those in $a_s P^{(0)}(\omega)$,
\begin{equation}
P^{{\rm SL}(0)}(\omega)=
\left( \begin{array}{cc}
0 & -3C_F \\
\frac{2}{3}T_R n_f & -\frac{11}{6}C_A-\frac {2}{3}T_R n_f
\end{array} \right),
\label{singlogsatLO}
\end{equation}
and apply the approximate result that follows from
the DLA at large $Q$,
\begin{equation}
D_{q,\overline{q}} =\frac{C_F}{C_A}D_g,
\label{DLArelforDquarkandDg}
\end{equation}
which can be derived, {\it e.g.}, from Eq.~(\ref{allDLinzindelPclosed}), the 
gluon component of Eq.~(\ref{DRAP}) becomes precisely the MLLA differential 
equation.
Therefore, we conclude that, since we do not use these two approximations, our
approach is more complete and accurate than the MLLA.

\section{Incorporation of hadron mass effects}
\label{sec:three}

We now incorporate hadron mass effects into our calculations, using a specific
choice of scaling variable.
For this purpose, it is helpful to work with light cone coordinates, 
in which any four-vector $V$ is written in the form 
$V=(V^+,V^-,{\mathbf V_T})$ with $V^{\pm}=(V^0 \pm V^3)/\sqrt{2}$ and
${\mathbf V_T}=(V^1,V^2)$.
In the center-of-mass (CM) frame, the momentum of the electroweak boson takes
the form
\begin{equation}
q=\left(\frac{\sqrt{s}}{\sqrt{2}},\frac{\sqrt{s}}{\sqrt{2}},
{\mathbf 0}\right).
\end{equation}
In the absence of hadron mass, $x_p$ (whose definition $x_p=2p/\sqrt{s}$
applies only in the CM frame) 
is identical to the light cone scaling variable
$\eta=p_h^+/q^+$. However, the definition $x_p=2p/\sqrt{s}$
applies only in the CM frame.
So, $\eta$ is a more convenient scaling variable for studying
hadron mass effects, since it is invariant with respect
to boosts along the direction of the hadron's spatial
momentum. Taking this direction to be the three-axis,
and introducing a mass $m_h$ for the hadron,
the momentum of the hadron in the CM frame reads
\begin{equation} 
p_h=\left(\frac{\eta \sqrt{s}}{\sqrt{2}},\frac{m_h^2}{\sqrt{2}\eta \sqrt{s}},
{\mathbf 0}\right).
\end{equation}
Therefore, the relation between the two scaling variables
in the presence of hadron mass is
\begin{equation}
x_p=\eta \left(1-\frac{m_h^2}{s\eta^2}\right).
\end{equation}
Note that these two variables are approximately equal
when $m_h \ll x_p \sqrt{s}$, {\it i.e.}\ hadron mass effects cannot be
neglected when $x_p$ (or $\eta$) are too small.

In the leading-twist component of the cross section after factorization, the
hadron $h$ is produced by fragmentation from a real, massless parton of
momentum
\begin{equation}
k=\left(\frac{p_h^+}{y},0,{\mathbf 0}\right).
\end{equation}
The $+$ component of everything other than this parton and of everything 
produced by the parton other than the observed hadron $h$ must be positive,
implying that $y \geq \eta$ and $y \leq 1$, respectively.
As a generalization of the massless case, we assume the 
cross section we have been calculating is $(d\sigma/d\eta)(\eta,s)$,
{\it i.e.}\ 
\begin{equation}
\frac{d\sigma}{d\eta}(\eta,s)=\int_{\eta}^1 \frac{dy}{y}\,
\frac{d\sigma}{dy}(y,s,Q^2)D\left(\frac{\eta}{y},Q^2\right),
\end{equation}
which is related to the measured observable $(d\sigma/dx_p)(x_p,s)$ via
\begin{equation}
\frac{d\sigma}{dx_p}(x_p,s)=\frac{1}
{1+m_h^2/[s\eta^2(x_p)]}\,
\frac{d\sigma}{d\eta}(\eta(x_p),s).
\end{equation}
Note that the effect of hadron mass is to reduce the size of the cross section
at small $x_p$ (or $\eta$).

The data we are studying are described by the sum of the production cross
sections for each light charged hadron species, being the charged pion, the
charged kaon, and the (anti)proton, whose masses (140, 494, and 938~MeV,
respectively) are substantially different.
Therefore, separate FFs and hadron masses for each of the three species are
needed.
Clearly, so many free parameters would not be constrained by these data. 
However, since most of the produced particles are pions, it is reasonable to 
take the final-state hadrons to have the same mass, so that in our approach 
it is reasonable to use a single hadron mass parameter $m_h$, 
whose fitted result should be closer to the pion mass than the proton mass, 
{\it i.e.}\ around 300~MeV.
Any significant deviation from this value would imply that some physics has
not been accounted for.

\section{Numerical analysis}
\label{sec:four}

In a first test of our approach, we compare its effects on fits of quark 
and gluon FFs to $e^+ e^-$ data with the standard FO DGLAP evolution.
We use normalized differential-cross-section data for the process
$e^+ e^- \rightarrow (\gamma,Z) \rightarrow h+X$, where $h$ is a light charged
hadron and $X$ is anything else, from TASSO at $\sqrt{s}=14$, 35, 44
\cite{Braunschweig:1990yd}, and 22~GeV \cite{Althoff:1983ew}, MARK~II
\cite{Petersen:1987bq} and TPC \cite{Aihara:1988su} at 29~GeV, TOPAZ at 58~GeV
\cite{Itoh:1994kb}, ALEPH \cite{Barate:1996fi}, DELPHI \cite{Abreu:1996na}, L3
\cite{Adeva:1991it}, OPAL \cite{Akrawy:1990ha}, and SLC \cite{Abrams:1989rz}
at 91~GeV, ALEPH \cite{Buskulic:1996tt} and OPAL \cite{Alexander:1996kh} at
133~GeV, DELPHI at 161~GeV \cite{Ackerstaff:1997kk}, and OPAL at 172, 183,
189 \cite{Abbiendi:1999sx}, and 202~GeV \cite{Abbiendi:2002mj}.
We place a small-$x$ cut \cite{Albino:2004yg} on our data of
\begin{equation}
\xi =\ln\frac{1}{x} < \ln \frac{\sqrt{s}}{2M},
\label{xicutwithm}
\end{equation}
where $M$ is a mass scale of $O(1$~GeV$)$.
We fit the gluon FF, $D_g(x,Q_0^2)$, as well as the quark FFs,
\begin{eqnarray}
D_{dsb}(x,Q_0^2)&=&\frac{1}{3}\left[D_d(x,Q_0^2)+D_s(x,Q_0^2)
\right.\nonumber\\
&&{}+\left.D_b(x,Q_0^2)\right],
\nonumber\\
D_{uc}(x,Q_0^2)&=&\frac{1}{2}\left[D_u(x,Q_0^2)+D_c(x,Q_0^2)\right],
\end{eqnarray}
where $Q_0=14$~GeV.
Since the hadron charge is summed over, we set $D_{\overline{q}}=D_q$.
For each of these three FFs, we choose the parameterization
\begin{equation}
D(x,Q_0^2)=N\exp(-c\ln^2 x)x^{\alpha} (1-x)^{\beta},
\label{genparam}
\end{equation}
which, at small $x$, is a Gaussian in $\xi$ for $c>0$ with center positive in
$\xi$ for $\alpha<0$ as is found to be the case,
while it reproduces the standard parameterization ({\it i.e.}\ that without
the  $\exp(-c\ln^2 x)$ factor) used in global fits at intermediate and large
$x$.
We use Eq.~(\ref{DLArelforDquarkandDg}) to motivate the simplification
\begin{equation}
c_{uc}=c_{dsb}=c_g,\qquad
\alpha_{uc}=\alpha_{dsb}=\alpha_g
\label{constraintsonac}
\end{equation}
to our parameterization. We also fit $\Lambda_{\rm QCD}$, so that there is a
total of 9 free parameters. Since we only use data for which
$\sqrt{s}>m_b$, where $m_b\approx 5$~GeV is the mass of the bottom
quark, and since $Q_0>m_b$, we take $n_f=5$.
While the precise choice for $n_f$ does not matter in the DLA, calculations in
the FO approach strongly depend on it. 

Since the theoretical error in our LO approach is larger than that at NLO,
some of our results should be interpreted somewhat qualitatively.
In particular, the result for $\Lambda_{\rm QCD}$ has a theoretical error of a
factor of $O(1)$.

Without the $(1-x)^{\beta}$ factors, Eq.~(\ref{DLArelforDquarkandDg}) implies
that
\begin{equation}
N_{uc}=N_{dsb}=\frac{C_F}{C_A} N_g.
\label{approxrelbetweenNs}
\end{equation}
Since the $(1-x)^\beta$ factors are important at large $x$, where
Eq.~(\ref{DLArelforDquarkandDg}) is no longer valid, we do not impose
Eq.~(\ref{constraintsonac}), but rather test its validity over the whole $x$
range.

\begin{table}[ht!]
\caption{\label{tab1}Parameter values for the FFs at $Q_0=14$~GeV
parameterized as in Eq.~(\ref{genparam}) from a fit, with
$\chi^2_{\rm DF}=3.0$, to all data listed in the text using DGLAP evolution in
the FO approach to LO.
The fit also yields $\Lambda_{\rm QCD}=388$~MeV.}
\begin{tabular}{ccccc}
\hline\hline
FF & $N$ & $\beta$ & $\alpha$ & $c$ \\
\hline
$g$ & 0.22 & $-0.43$ & $-2.38$ & 0.25 \\
$u+c$ & 0.49 & 2.30 & [$-2.38$] & [0.25] \\
$d+s+b$ & 0.37 & 1.49 & [$-2.38$] & [0.25] \\  
\hline\hline
\end{tabular}
\end{table}
\begin{figure}[ht!]
\includegraphics*[width=0.48\textwidth]{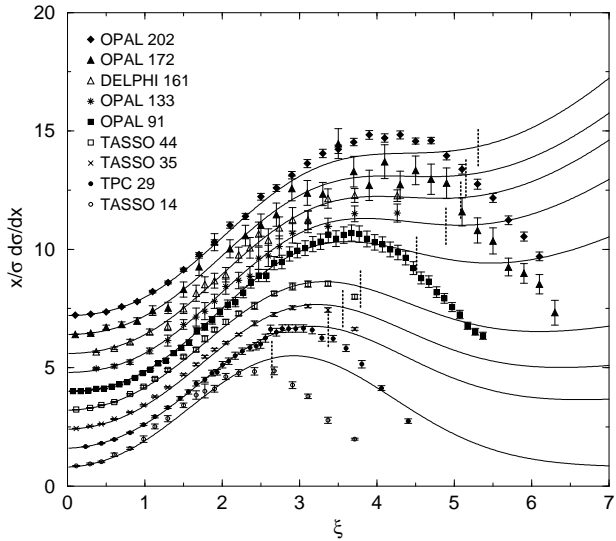}
\caption{\label{fig1}Fit to data as described in Table~\ref{tab1}.
Some of the data sets used for the fit are shown, together with their
theoretical predictions from the results of the fit.
Data to the right of the vertical dotted lines are not used.
Each curve is shifted up by 0.8 for clarity.}
\end{figure}
We first perform a fit to all data sets listed above 
using standard LO DGLAP evolution, {\it i.e.}\ Eq.~(\ref{DGLAPx})
without the replacement in Eq.~(\ref{replace}).
We fit to those data for which Eq.~(\ref{xicutwithm}) is obeyed with
$M=0.5$~GeV. 
This gives a total of 425 data points out of the available 492.
We obtain $\chi^2_{\rm DF}=3.0$ (or 2.1 after subtraction of the contribution
to $\chi^2$ from the TOPAZ data, which is the only data set whose individual
$\chi^2_{\rm DF}$ is greater than 6), and the results are shown in
Table~\ref{tab1} and Fig.~\ref{fig1}. 
It is clear that FO DGLAP evolution fails in the description of the peak
region and shows a different trend outside the fit range.
The $\exp(-c\ln^2 x)$ factor does at least allow for the fit range to be
extended to $x$ values below that of $x=0.1$, the lower limit of most global
fits, to around $x=0.05$ ($\xi=3$) for data at the larger $\sqrt{s}$ values.
Note that $\beta_g$ is negative, while kinematics require it to be positive.
However, this clearly does not make any noticeable difference to the cross
section.

\begin{table}[ht!]
\caption{\label{tab2}Parameter values for the FFs at $Q_0=14$~GeV
parameterized as in Eq.~(\ref{genparam}) from a fit, with
$\chi^2_{\rm DF}=2.1$, to all data listed in the text using DGLAP evolution in
the  DLA-improved approach to LO.
The fit also yields $\Lambda_{\rm QCD}=801$~MeV.}
\begin{tabular}{ccccc}
\hline\hline
FF & $N$ & $\beta$ & $\alpha$ & $c$ \\
\hline
$g$ & 1.60 & 5.01 & $-2.63$ & 0.35 \\
$u+c$ & 0.39 & 1.46 & [$-2.63$] & [0.35] \\
$d+s+b$ & 0.34 & 1.49 & [$-2.63$] & [0.35] \\  
\hline\hline
\end{tabular}
\end{table}
\begin{figure}[ht!]
\includegraphics*[width=0.48\textwidth]{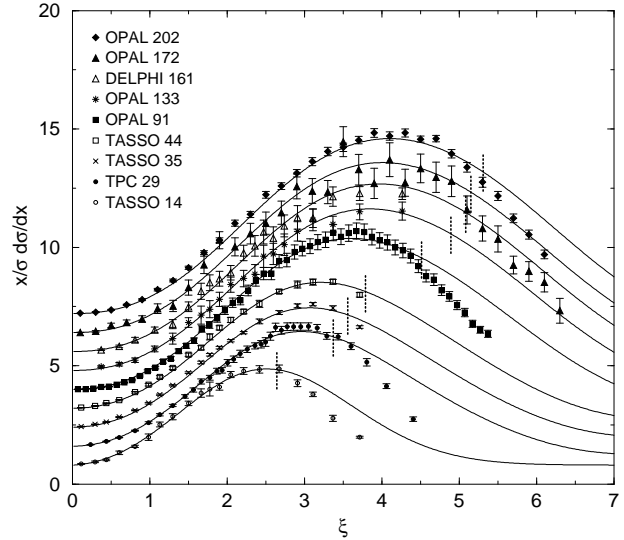}
\caption{\label{fig2}Fit to data as described in Table~\ref{tab2}.}
\end{figure}
Now we perform the same fit again, but using our approach, {\it i.e.}\
Eq.~(\ref{DGLAPx})
with the replacement in Eq.~(\ref{replace}), for the evolution.
The results are shown in Table~\ref{tab2} and Fig.~\ref{fig2}. 
We obtain $\chi^2_{\rm DF}=2.1$ (or 1.4 without the TOPAZ data, the individual
$\chi^2_{\rm DF}$ for each remaining data set being less than 3), a
significant improvement relative to the fit above with FO DGLAP evolution.
This should also be compared with the fit to the same data in
Ref.~\cite{Albino:2004xa}, where DL resummation
is used within the MLLA but with neither FO terms nor quark freedom 
({\it i.e.}\ Eq.~(\ref{DLArelforDquarkandDg}) is imposed over the
whole $x$ range) and $\chi^2_{\rm DF}=4.0$ is obtained. 
The data around the peak is now much better described.
The energy dependence is well reproduced up to the largest $\sqrt{s}$ value,
202~GeV.
At $\sqrt{s}=14$~GeV, the low-$x$ description of the data is extended from
$x=0.1$ in the unresummed case down to 0.06 in the resummed case, and from
$x=0.05$ to 0.005 at $\sqrt{s}=202$~GeV.

We conclude that, relative to the MLLA, the FO contributions in the evolution,
together with freedom from the constraint of
Eq.~(\ref{DLArelforDquarkandDg}), make a significant improvement to the
description of the data for $\xi$ from zero to just beyond the peak. 
The value $\Lambda_{\rm QCD}\approx 800$~MeV is somewhat larger than the 
value 480~MeV, which we obtain from a DGLAP fit in the large-$x$ range
($x>0.1$).
Had we made the usual DLA (MLLA) choice $Q=\sqrt{s}/2$ instead of our choice
$Q=\sqrt{s}$ as is done in analyses using the DGLAP equation, we would have
obtained half this value for $\Lambda_{\rm QCD}$.
In an analysis at NLO, the choice of $Q$ is less relevant, since the
theoretical error on $\Lambda_{\rm QCD}$ is smaller.
In addition, while the DL resummation greatly improves the description around
the peak, it is still not perfect.
$N_g$ exceeds the value expected from Eq.~(\ref{DLArelforDquarkandDg}) 
by a factor of about 2. However, note that $N_g$ is weakly constrained,
since the gluon FF couples to the data only through the evolution, requiring
gluon data to be properly constrained.
These problems are related to the worsening of the description of the data on
moving beyond the peak, since fits in which the cuts are moved to larger
$\xi$ values give an increase in $\Lambda_{\rm QCD}$ and $N_g$, as well as in
$\chi^2_{\rm DF}$.

\begin{figure}[ht!]
\includegraphics*[width=0.48\textwidth]{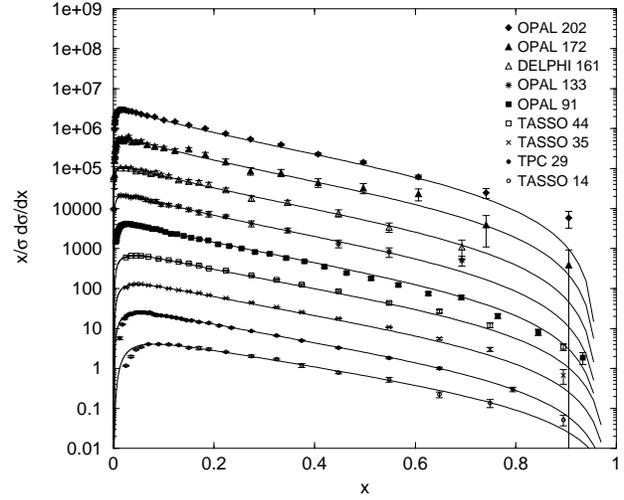}
\caption{\label{fig3}As in Fig.~\ref{fig2}, but with the cross section on a
logarithmic scale versus $x$.
Each curve, apart from the lowest one, has been rescaled relative to the one
immediately below it by a factor of 5 for clarity.}
\end{figure}
Figure~\ref{fig2} is repeated in Fig.~\ref{fig3}, to show more clearly the
good quality of the fit at intermediate and large $x$.
A couple of points at $x=0.9$ are not well described, although the data here
are scarce and have larger errors.
This shows that the DL resummation does not deteriorate the quality of the FO
description at large $x$.
This is further confirmed by noting that, for the above fit in our approach,
the subsample of the data at $x>0.1$ yields $\chi^2_{\rm DF}=3.0$, which is
close to the result $\chi^2_{\rm DF}=2.9$ obtained from fitting to the same
large-$x$ data using only the FO approach. 

\begin{table}[ht!]
\caption{\label{tab3}Parameter values for the FFs at $Q_0=14$~GeV
parameterized as in Eq.~(\ref{genparam}) from a fit, with
$\chi^2_{\rm DF}=2.0$, to all data listed in the text using DGLAP evolution in
the DLA-improved approach to LO incorporating hadron mass effects.
The fit also yields $\Lambda_{\rm QCD}=399$~MeV and $m_h=252$~MeV.}
\begin{tabular}{ccccc}
\hline\hline
FF& $N$ & $\beta$ & $\alpha$ & $c$ \\
\hline
$g$ & 1.59 & 7.80 & $-2.65$ & 0.33 \\
$u+c$ & 0.62 & 1.43 & [$-2.65$] & [0.33] \\
$d+s+b$ & 0.74 & 1.60 & [$-2.65$] & [0.33] \\ 
\hline\hline
\end{tabular}
\end{table}
\begin{figure}[ht!]
\includegraphics*[width=0.48\textwidth]{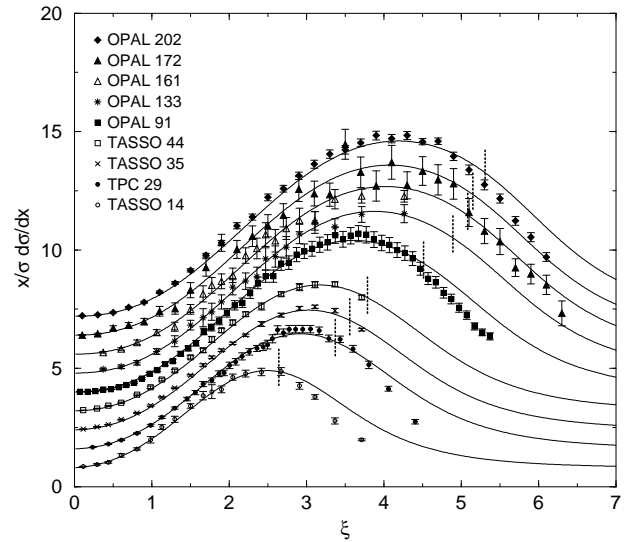}
\caption{\label{fig4}Fit to data as described in Table~\ref{tab3}.}
\end{figure}
We now perform the last fit again, but with $m_h$ included in the list of
free parameters.
We obtain the results in Table~\ref{tab3} and Fig.~\ref{fig4}, and the values
$\Lambda_{\rm QCD}=399$~MeV and $m_h=252$~MeV.
Treatment of hadron mass effects renders the value of $\Lambda_{\rm QCD}$
obtained in the fit with DL resummation more reasonable and slightly improves
the latter, yielding $\chi^2_{\rm DF}=2.0$.
We conclude, therefore, that to improve the large-$\xi$ description and to
achieve a reasonable value for $\Lambda_{\rm QCD}$, both DL resummation and
treatment of hadron mass effects are required.

\section{Conclusions}
\label{sec:five}

In conclusion, we have proposed a single unified scheme which can describe a
larger range in $x$ than either FO DGLAP evolution or the DLA, by implementing
SGL resummation in the standard DGLAP formalism in a minimal and unambiguous
way through any order of perturbation theory.
Our present limitation to LO is caused by the lack of knowledge on SGL
resummation necessary to match NLO DGLAP evolution.
The actual inclusion of higher orders, although conceptually straightforward
in our scheme, requires a separate analysis, which lies beyond the scope of
Ref.~\cite{Albino:2005gg} and is left for future work.
Further improvement in the small-$x$ region can be expected from the inclusion
of resummed SLs.
Our scheme allows for a determination of quark and gluon FFs over a wider
range of data than previously achieved, and should be incorporated into global
fits of FFs such as that in Ref.~\cite{Albino:2005me}, since the current
range of $0.1<x<1$ is very limited.

Since FFs and their DGLAP evolution are universal, our approach should be
expected to also improve the description of inclusive hadron production from
reactions other than $e^+ e^-$ annihilation, such as those involving at least
one proton in the initial state.

\bigskip 
\begin{acknowledgments}
The author is grateful to the organizers of the IPM School and Conference on
Lepton and Hadron Physics for the kind invitation
and for creating such a stimulating atmosphere, and to Simon Albino,
Gustav Kramer, and Wolfgang Ochs for their collaboration on the work presented
here.
This work was supported in part by the Deutsche Forschungsgemeinschaft     
through Grant No.\ KN~365/3-1 and by the Bundesministerium f\"ur Bildung und  
Forschung through Grant No.\ 05~HT4GUA/4.
\end{acknowledgments}

\bigskip 

\end{document}